# Method for the Monte Carlo Simulation of Lipid Monolayers including Lipid Movement

J. Griesbauer[1,2], H. Seeger [3], A. Wixforth[1], M.F. Schneider[2]

[1] *University of Augsburg, Experimental Physics I, D-86159 Augsburg, Germany*
[2] *Boston University, Dept. of Mechanical Engineering, Boston-Massachusetts, USA*
[3] *Centro S3, CNR-Istituto di Nanoscienze, Via Campi 213/A, 41125 Modena, Italy*

**Abstract**

A two-state-model consisting of hexagonally connected lipids being either in the ordered or disordered state is used to set up a Monte Carlo Simulation for lipid monolayers. The connection of the lipids is realized by Newtonian springs emulating the surfaces elasticity and allowing for the calculation of translational movement of the lipids, whereas all necessary simulation parameters follow from experiments. Simulated monolayer isotherms can be directly compared to measured ones concurrently allowing the calculation of the experimentally hardly accessible monolayer heat capacity.

**Introduction**

Monte Carlo (MC) simulations of the lipid main phase transition in bi- and monolayers using simple two-state Ising like models [1][2][3] were successfully utilized for a variety of applications. Among the latter are the simulation of two-component lipid membranes [4][5], the investigation of lipid-protein interactions [6][7], the interpretation and reproduction of FSC, FTIR or DSC measurements [8][9] and the calculation of lipid monolayer isotherms [10][11]. All of these simulations are based on static hexagonal lattice models not allowing for lipid movement.
We use the core of these two-state Ising like models to implement the lipid ordered and disordered states [1][2][3][12][13] and embed this base into a flexible hexagonal lattice established by Newtonian springs as it is shown in the picture inset of fig. 1. During the cycles of a simulation, the thermodynamic potential used for the MC core gets completed by the spring energies, whereas lipid movement is done in a separate step according to the Hookean and Newtonian laws. Eventually the experimental fixation of the simulation parameters, enables us to reproduce DPPC monolayer isotherms indicating details like ordered lipid domains during phase transition, along with the possibility to access the experimentally hardly measureable heat capacity profile of a lipid monolayer.

**Materials and Methods**

Lipid 1,2-Dipalmitoyl-sn-Glycero-3-Phosphocholine (DPPC) dissolved in chloroform was purchased from Avanti Polar Lipids (*Birmingham, Al. USA*) and used without further purification.

All pressure-isotherm measurements were done on a standard film balance (*NIMA, Coventry, England*) connected to a heat bath allowing for temperature regulation. Standard isotherms of DPPC monolayers on pure water (*18 MΩ/cm*) were recorded and used.

**Basic Theory**

The study at hand is based on the two-state Ising like model, which was applied beforehand [1][2][3][12][13] and which shall be briefly described in the following.

Hexagonally arranged lipids are considered to occur either in the energetic lower, ordered state (gel like) or the energetic higher, disordered state (fluid like) and to only experience interactions with the six nearest neighbours (for definition a monolayer consisting of lipids in the ordered state is in the liquid condensed phase, one consisting of lipids in the disordered state is in the liquid expanded phase).

In the run of a simulation a Markov chain is generated using the Glauber algorithm by randomly selecting a lipid in state *s* and calculating its transition probability $P(s \to s')$ for changing to the opposing state $s'$:

$$P(s \to s') = \frac{K}{1+K} \quad (1),$$

with

$$K = e^{\frac{-\Delta G}{RT}} \quad (2),$$

where *T* is the temperature, *R* the general gas constant and *ΔG* the change in the Gibbs free energy for a change of state of the selected lipid. The latter's state *s* is changed to *s'*, when a random number $z_1 \in [0;1]$ is smaller than $P(s \to s')$, or remaines *s* if not. Thereby a new state of the system is defined and by successive iteration of the so defined MC-step a series of states (Markov chain) is produced. For this process the Gibbs free energy *ΔG* of a state change is necessary and calculated by

$$\Delta G = \Delta H - T \Delta S + \Delta n_{gf} \omega_{gf} \quad (3),$$

where *ΔH* and *ΔS* are the enthalphy and entropie difference due to a ordered-disordered state change. It is $\Delta S \approx \frac{\Delta Q}{T_m}$ (or $\Delta S \approx \frac{\Delta H}{T_m}$ for an isobaric process), where $T_m$ is the phase transition temperature of a lipid (at zero lateral pressure in a monolayer). $\Delta n_{fg}$ equals the number of changes of nearest neighbours being in the opposing state to the selected lipid, whereas $\omega_{gf}$ is the interaction or cooperative parameter between lipids of different states. In this procedure the necessary parameters *ΔH*, $T_m$ and $\omega_{gf}$ are well known for pure DPPC [1][2][3][13].

## Results

### Modification of the Basic Theory

For the work at hand we assume the lipid lattice to be established by springs hexagonally interconnecting the lipids as it is shown in the picture inset of fig. 1 (it shall be mentioned here, that the hexagonal structure is a good approximation for the ordered state [14], but only a rough assumption for the disordered state). Each lipid gets its own set of six springs having resting lengths and spring constants according to the area and compressibility of the lipid. Those springs introduce an energy into the system, which mimics the surface energy of the lipid membrane and which needs to be considered in terms of equ. (3). This is done by using the thermodynamic relation $\Delta H = \Delta Q - A\Delta \Pi$ resulting in:

$$\Delta G = \Delta Q (1 - \frac{T}{T_m}) + \Delta n_{gf} \omega_{gf} + A \Delta \Pi \quad (4).$$

The value of $A\Delta\Pi$ in equ. (4) equals the change of surface energy or (for the model at hand) spring energy generated due to the state change of a selected lipid. The difference arises due to the change of resting lengths (area) and spring constants (compressibility) of the springs, when the selected lipid changes its state (from ordered to disordered or from disordered to ordered). To begin with, the spring energy change $\Delta E_{Sp}$ of all involved springs (six springs of the selected lipid each connected to one spring belonging to a neighbouring lipid resulting in twelve springs all together) is calculated for a state change of a selected lipid *1*. For each set of two springs connecting lipid *1* (spring constants $k_{1:changed/unchanged}$, determined later) with its neighbours $i$ (spring constant $k_i$, determined later), this is done by assuming all springs to be stretched according to their combined spring constant. This is done for the *changed* and *unchanged* state of lipid *1* yielding the expressions $k_{changed/unchanged} = (1/k_{1:changed/unchanged} + 1/k_i)^{-1}$. The resting lengths of the involved springs of lipid *1* and $i$, shall be defined by $x_{01changed/unchanged}$ and $x_{0i}$, whereas the distance between the lipid positions is given by $x_{1-i}$. Using these definitions, $\Delta E_{Sp}$ is calculated by the following terms:

$$\Delta E_{1-i} = \frac{1}{2} k_{unchanged} (x_{1-i} - x_{01unchanged} - x_{i0})^2 - \frac{1}{2} k_{changed} (x_{1-i} - x_{01changed} - x_{i0})^2$$

$$\Delta E_{Sp} = \sum_{i: nearest\ neighbours} \Delta E_{1-i}$$

(5)

This $\Delta E_{Sp}$ comprises the relaxation of lipid *1 and all of its neighbours*, thereby approximatelly associating half of $\Delta E_{Sp}$ to lipid *1*. Because only the selected lipid *1* may change its state and therefore relax its spring energies, it is assumed that $A\Delta\Pi \approx \frac{\Delta E_{Sp}}{2}$, resulting in the final expression for $\Delta G$.

$$\Delta G = \Delta Q(1 - \frac{T}{T_m}) + \Delta n_{gf} \omega_{gf} + \frac{\Delta E_{Sp}}{2} \quad (6)$$

For the evaluation of equ. (5) the relationship of spring resting lengths ($x_0$) and lipid areas ($A_0$) is needed, whereas the two possible states (ordered: *g*, disordered: *f*) imply two possible lipid areas

$A_{0f/g}$ and therefore resting lengths $x_{0f/g}$:

$$A_{0f/g} = \frac{3\sqrt{3}\, x_{0f/g}^2}{2} \Leftrightarrow x_{0f/g} = \sqrt{\frac{2 A_{0f/g}}{3\sqrt{3}}} \qquad (7)$$

Eventually the spring constants $k_1$ and $k_i$ of lipid *1* and its neighbours *i* in equ. (5) have to be determined using the compressibilities $\kappa_l$ and $\kappa_i$ of the lipids, which can have two different values $\kappa_f$ and $\kappa_g$ for the ordered (*g*) and disordered (*f*) state. Using the resting lengths $x_{01}$ and $x_{0i}$ and the distance $x_{1-i}$ of lipid *1* and *i* Appendix A derives the following expressions for the spring constants.

$$\Delta x_{1-i} = x_{1-i} - x_{01} - x_{i0}$$

$$k_j = \frac{\sqrt{3}}{\kappa_j}\left(1 + \frac{1}{x_{0j}}\Delta x_{1-i} + \frac{1}{4 x_{0j}^2}\Delta x_{1-i}^2\right) \quad \text{with } j = 1, i \qquad (8)$$

Finally, by summing over all changes of the spring energies, the complete spring energy $E_{Sp}$ of the entire system is calculated and used to evaluate the surface pressure $\Pi$ using the expression $\Pi A = E_{sp}$. At the same time summing over all $\Delta G$ defines the complete Gibbs free energy $G$ (and thereby enthalpy $H$) of the entire system.

*Extension of the model by lipid movement*

The described calculus for $\Delta G$ (*Modification of the Basic Theory*) is used to determine the probability of a state change of a randomly selected lipid and thereby defines a MC-step of the simulation. For the simulations at hand we additionally introduce a movement step, which moves the lipids according to the spring forces. This movement does not change the hexagonal network set up by the springs, but relaxes the springs towards their energetic minimum.

First of all, the force $\vec{F} = \sum_{i:\, nearest\, neighbours} \vec{n}_{1-i}\, k_{comb}\, \Delta x_{1-i}$ on a randomly selected lipid *1* is calculated using the distances $\Delta x_{1-i}$ (see equ. (8)), normalized vectors $\vec{n}_{1-i}$ pointing to the neighbouring lipids *i* and the combined spring constants $k_{comb} = (1/k_1 + 1/k_i)^{-1}$. By introducing an arbitrary time constant $\delta$ and friction constant $c_f$ the new position $\vec{x}_{new}$ and velocity $\vec{v}_{new}$ of the selected lipid (mass *m*) is evaluated, based on its initial position $\vec{x}_{old}$ and velocity $\vec{v}_{old}$ at the beginning of the movement step:

$$\vec{a} = \frac{\vec{F}}{m} - c_f \vec{v}_{old}$$
$$\vec{v}_{new} = \vec{v}_{old} + \vec{a}\,\delta \qquad (9)$$
$$\vec{x}_{new} = \vec{x}_{old} + \vec{v}_{old}\,\delta + \frac{1}{2}\vec{a}\,\delta^2$$

For the cycle of a simulation this movement step and the beforehand defined MC-step get executed based on a randomly chosen number $z_2 \in [0;1]$: for the randomly selected lipid the movement step is executed, when $z_2 < 0.5$, whereas for $z_2 > 0.5$ the MC-step is executed. Fig. 2 shows the entire cycle including both possible steps.

*Determination of lipid parameters*

Before the first simulations can be processed, the lipid parameters $\kappa_f$, $\kappa_g$, $x_{0f}$, $x_{0g}$, $T_m$, $\Delta H$ and $\omega_{gf}$ have to be determined. For DPPC this was done using single measurements fixing those parameters to constant values.

- $x_{0f}$ and $x_{0g}$ were obtained using the lipid areas $A_g$ and $A_f$ of the ordered (*g*) and disordered (*f*) states as they are implied by the isotherms of a DPPC monolayer at two different temperatures of *3°C* and *42°C*. At these temperatures the monolayer lipids are approximately either in the ordered state (*3°C*) or the disordered state (*42°C*) at all times, so that the areas of the first pressure rises of the isotherms imply $A_f \approx 90\text{Å}$ (*42°C*) or $A_g \approx 52\text{Å}$ (*3°C*) as it is shown in fig. 1. X-Ray Reflectivity measurements [15] propose $A_g$ of DPPC to be about *48Å*, which is in good agreement with our measurements. Since the chosen $A_f$ gets additionally influenced by free volume effects [10] (hydrophobic chains lying on the water surface) a reasonable comparison of $A_f$ to literature values is not attempted.
- $\kappa_f$ and $\kappa_g$ can be extracted from the compressibility curves, which are shown in the inset of fig. 1 and got extracted in terms of the expression $\kappa_T = \frac{-1}{A}\left(\frac{\partial A}{\partial \Pi}\right)_T$ by using the isotherm curves *Π(A)*. Since constant values for $\kappa_f$ and $\kappa_g$ are assumed, the minimas of the experimental compressibility curves were used for the determination, resulting in $\kappa_f \approx 10 m/N$ and $\kappa_g \approx 4 m/N$.
- $T_m$ is defined by the transition temperature of the ordered to the disordered state for a lipid monolayer. Therefore it can be extracted by lowering the temperature during several isotherms of a DPPC monolayer, until that temperature ($T_m$) is reached, where the phase transition plateau in the isotherms vanishes, indicating, that all lipids remain in the ordered state over the whole isotherm for temperatures lower than $T_m$. From our experiments we obtain a value of $T_m \approx 14°C$, which is in good agreement with estimations of around *15°C* made in [1].
- The constants $\Delta Q \approx \Delta H \approx 36700 J/mol$ and $\omega_{gf} \approx 1187 J/mol$ are adjusted and adapted from earlier Monte Carlo Simulations (of the type described in *Basic Theory*) made in [6][7].

*Choice of the constant δ and $c_f$*

The time constant *δ* is an arbitrary value discretizing the lipid movement to fit the cycle wise simulation model. In this sense, the friction constant $c_f$ not only accounts for natural friction, but also avoids the increase of spring energy due to the discretization. Since the model at hand does not intend to comprehensively consider viscous friction effects or the like, $c_f$ may be consequently chosen as small as possible.
As the following discussion will show, the simulation results depend only little on the actual value of *δ* or $c_f$ as long as both values are chosen in a reasonable range, which i.e. prevents, that the discrete lipid movements get comparable to the order of the lattice constant itself (rendering a stable simulation literally impossible). Nevertheless *δ* and $c_f$ were characterized using a set of simulations on a *32x31* lattice at a runtime of *30K* cycles per lipid (at *24-°C*) for different pairs of *δ* and $c_f$ (*30K* cycles per lipid produce reasonable results as discussed in *Lattize Size Effects and Stability*). An example for a resulting isotherm of such a simulation run is shown together with the accordant measurement for a DPPC monolayer in fig. 3 a). For the further analysis of the simulation behaviour dependant on *δ* and $c_f$, the integrated difference between measurement and simulation is used as additionally outlined in fig. 3.a). Rows of those differences for three different values of $c_f$ over a wide range of $\delta$ are presented in fig. 3 b). Seemingly for the smallest value of $c_f$ the steepest course of integrated differences occurs, rendering $c_f=2\cdot 10^{-15}$ as a not reasonable choice due to the relatively large influence of changes in $\delta$ on the simulation results. Contrastly $c_f=4\cdot 10^{-15}$ and $c_f=6\cdot 10^{-15}$ produce flat curves for the row of values for $\delta$, so that the smaller value $c_f=4\cdot 10^{-15}$ is selected for further use. In the course of values for $\delta$ at $c_f=4\cdot 10^{-15}$, $\delta=0$ (lipid movement is eliminated) and $\delta>4\cdot 10^{-14}$ (the integrated differences curve gets steeper again) mark the maximal and minimal reasonable values for $\delta$. Hence, an arbitrary value of $\delta=2.4\cdot 10^{-14}$, which lies in between those two limiting values and is in a flat region of the integrated differences curve, is chosen for the following simulations.

As this discussion implies, it is not intended to choose the optimal values for $\delta$ or $c_f$ here, especially in respect to the fact, that the rows in fig. 3 b) clearly show, that small differences in $\delta$ or $c_f$ would produce even smaller changes in the simulation results. Nevertheless the choice of $\delta$ and $c_f$ here, is of that kind, to prevent the model system from suffering under to large or destabilizing lipid movements (i.e. over whole lattice sites).

*Lattice Size Effects and Stability*

In the simulations at hand lattice sizes of *32x31* and *102x101* (*lipidsxlipids*) and a run time of minimal *30K* simulation cycles per lipid are used. For the further characterization of influences of runtime and lattice size simulations over *10K* to *100K* cycles were executed and show, that for runtimes larger than *30K* cycles per lipid no alteration of the simulation results are indicated (considering both isotherms and structural details). In detail, statistical quantities (calculated using thermodynamic averages as in equ. 10), increased their stability from *10K* to *20K* cycles, starting to converge for runtimes larger than *20K* cycles, whereas values for the lateral pressure and isotherms readily converged at runtimes of *10K* cycles.

At the same time, changing the size of the model system from *32x32* to *102x101* has influence on the shape of the simulated isotherms. Those influences are indicated at the beginning of the phase transition plateau, where the differences of simulated and measured isotherms differ. Simulated isotherms a slightly lower than measured ones, whereas this difference is positively altered by the lattice size, meaning that the bigger the lattice the less difference occurs. Simulations on a much larger lattice (*202x201*) were done and did not provide substantial improvement. Hence, for a first approach, it was refrained to further work on larger lattices due to the heavily increased computational runtimes.

*Comparison of measured and simulated isotherms*

To test for the feasibility and reasonability of the proposed model, we performed simulations of monolayer isotherms at different temperatures and compared them to the according experimental results. Fig. 4 a) shows both measured and simulated isotherms for three different temperatures at *20°C, 24°C* and *28°C*. Rather good agreement is revealed, demonstrated by comparable lengths of phase transition plateaus, whereas even the experimentally expected [16] linear shift of the plateau heights with temperature can be observed.

Considering the lattice structure itself, another phenomenological fact, as seen in fluorescent or Brewster angle microscopy studies [17][18][19], might be indicated by the simulations and is shown in fig. 4 b). During phase transition, persistent ordered domains evolve, which alter their general shape depending on temperature. Since our model is very minimalistic and doesn't account for details like lipid charges or chirality, this similarity has to be considered as a clue towards experiment and not as a detailed reproduction of domain growth and behaviour.

*Evaluation of the heat capacity of the simulated lipid monolayers*

As an example for an additionally accessible result, which is inherent to most MC type simulations, the calculation of the experimentally hardly accessible heat capacity $c_p$ of the monolayers is performed, using the fluctuation dissipation-theorem. This is done in terms of the thermodynamic averages $\langle H \rangle$ and $\langle H^2 \rangle$ of the ethalpy H, which are extracted and recorded during simulations:

$$c_p = \frac{\langle H^2 \rangle - \langle H \rangle^2}{RT} \quad (10)$$

The determined heat capacity curves $c_p$ along with the corresponding simulated compressibility curves $\kappa_T$ at the three different temperatures of *20°C, 24°C* and *28°C* are shown in fig. 5 (again $\kappa_T$ is calculated using the formula $\kappa_T = \frac{-1}{A}\left(\frac{\partial A}{\partial \Pi}\right)_T$ ). Several details are revealed. Firstly, as expected from the corresponding isotherms, the compressibility $\kappa_T$ shows a split maximum due to the form of the phase transition plateau (fig. 4 a) ). Secondly it can be seen, that $c_p$ and $\kappa_T$ have their average maxima at the same lateral pressures with a contrary course of peak heights with temperature ($c_p$ peaks increase with temperature, $\kappa_T$ peaks decrease with temperature). Indeed the coincidence of the lateral pressures of the maxima in $c_p$ and $\kappa_T$ has been proposed beforehand [16][20] using theoretical and experimental argumentations. Our simulations support this coincidence, whereas the detailed courses of $c_p$ and $\kappa_T$ indicate a more complex connection than the proposed direct proportionality in [16][20].

**Conclusion**

A simple Monte Carlo type method for the simulation of lipid monolayers (or bilayers) is introduced, which includes lipid movement and therefore is capable of mimicking the elastic behaviour of lipid membranes. The model consist of hexagonally arranged lipids interconnected by springs, whereas the spring energies get incorporated into the thermodynamic potential of the system and lipids are moved according to the spring forces in discrete steps.
The feasibility of the proposed model is tested by comparison of measured and simulated isotherms, revealing good agreement, even indicating structural parallels (domain growths during isotherms). Additionally the experimentally hardly accessible heat capacity profile of the simulated monolayers is determined supporting the coincidence of compressibility and heat capacity maxima during phase transition.
The proposed fairly simple simulation model tries to incorporate simple monolayer dynamics into static Monte Carlo type simulations, whereas a common home computer is capable of the computational requirements. Eventually this would make it possible to run larger simulations, or to detail the model by considering lipid charges, chirality or lipid diffusion. Even structural lipid systems like vesicles or dynamics like wave propagation on lipid monolayers [21] are easily feasible using or extending the proposed model.

**Appendix A**

When a lipid increases its area from $A_0$ by $\Delta A$ and has the compressibility $\kappa$ the energy $\frac{1}{2\kappa}\frac{\Delta A^2}{A_0}$ is needed. In the model at hand this energy has to equal the energy produced by the six springs of a lipid. Assuming that the lipid's springs with spring constant $k$ are stretched equally from $x_0$ by $\Delta x$, this means, that the energy $3k\Delta x^2$ has to be equal to $\frac{1}{2\kappa}\frac{\Delta A^2}{A_0}$. Firstly, $\Delta A$ and $\Delta x$ can be connected in the following way:

$$A_0 = \frac{3\sqrt{3}}{2}x_0^2, \; A_0 + \Delta A = \frac{3\sqrt{3}}{2}(x_0+\Delta x)^2 \Rightarrow \Delta A = \frac{3\sqrt{3}}{2}(2x_0\Delta x + \Delta x^2)$$

Equallizing $3k\Delta x^2$ and $\frac{1}{2\kappa}\frac{\Delta A^2}{A_0}$ yields:

$$3k\Delta x^2 = \frac{1}{2\kappa}\left(\frac{\left(\frac{3\sqrt{3}}{2}(2x_0\Delta x+\Delta x^2)\right)^2}{\frac{3\sqrt{3}}{2}x_0^2}\right) = \frac{3\sqrt{3}}{\kappa}\Delta x^2\left(1+\frac{\Delta x}{x_0}+\frac{\Delta x^2}{4x_0^2}\right)$$

$$\Rightarrow k = \frac{\sqrt{3}}{\kappa}\left(1+\frac{\Delta x}{x_0}+\frac{\Delta x^2}{4x_0^2}\right)$$

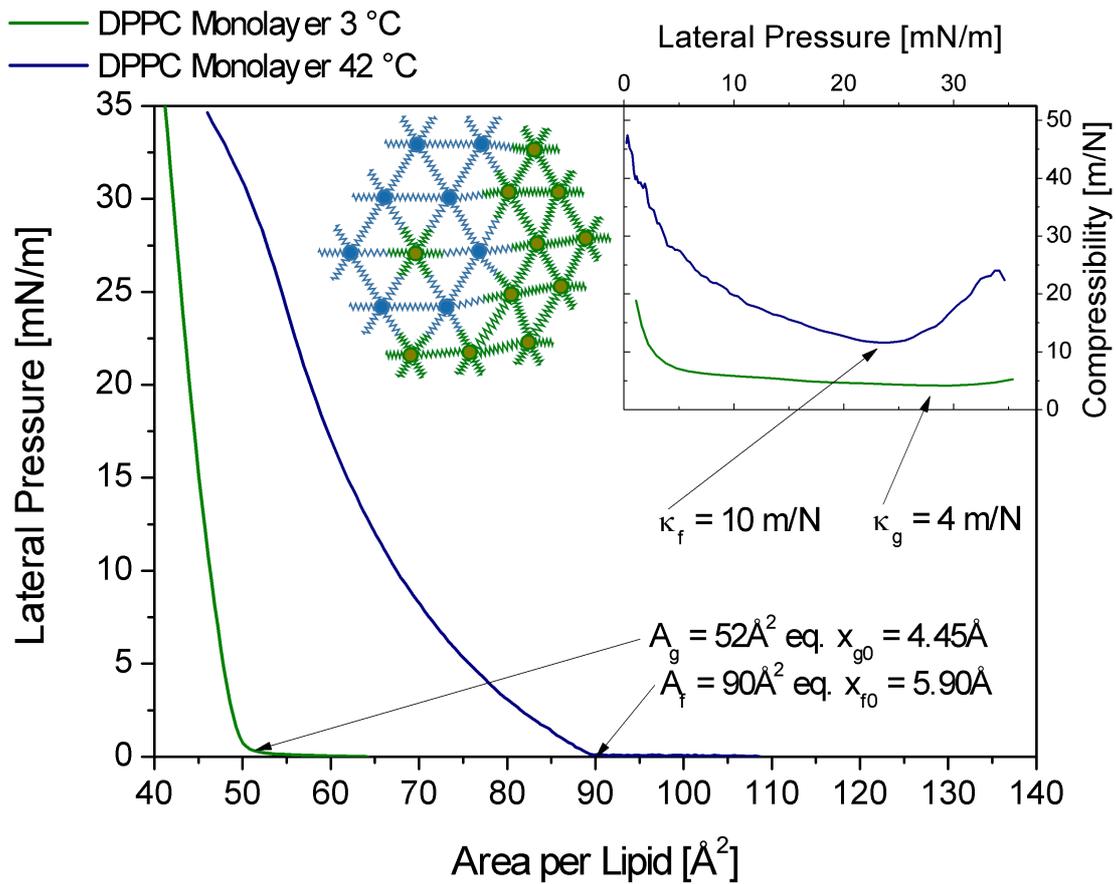

**Figure 1**

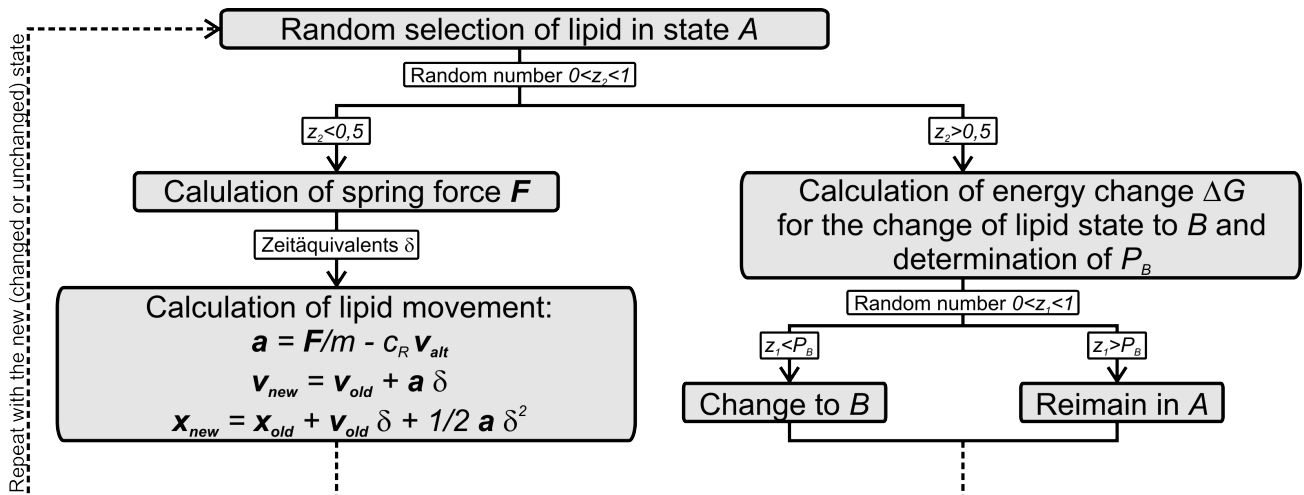

**Figure 2**

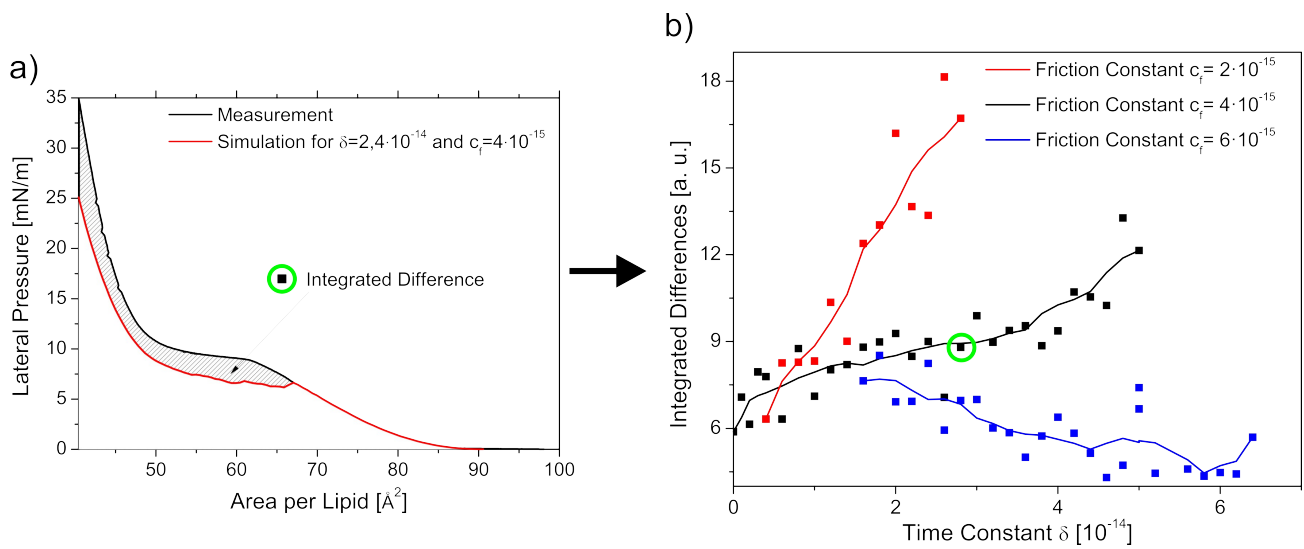

**Figure 3**

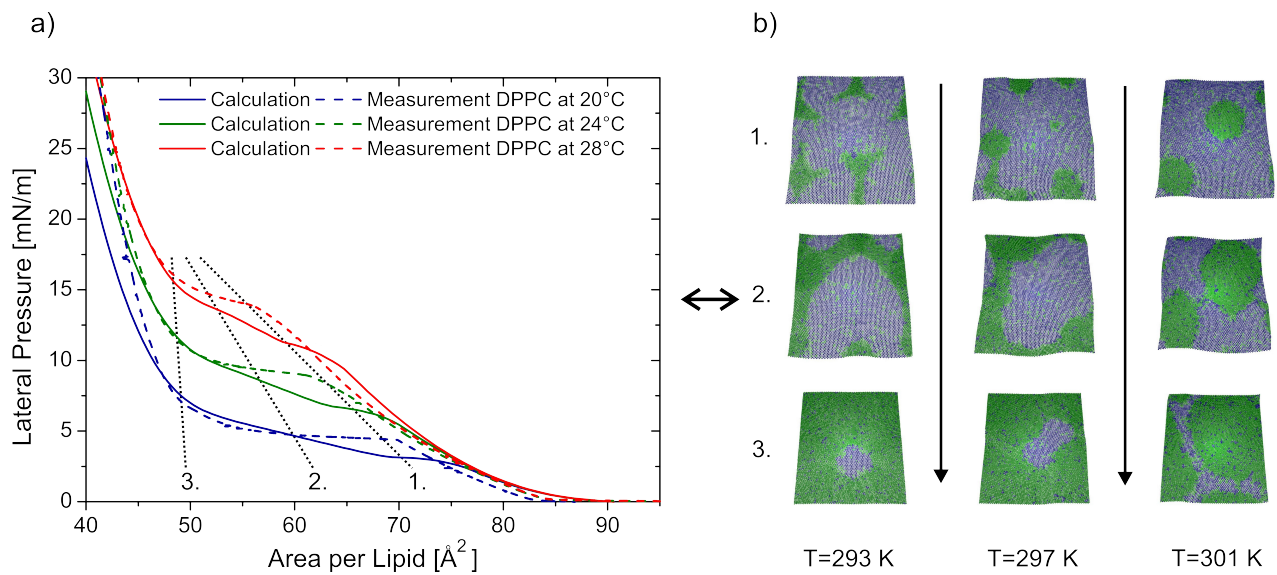

**Figure 4**

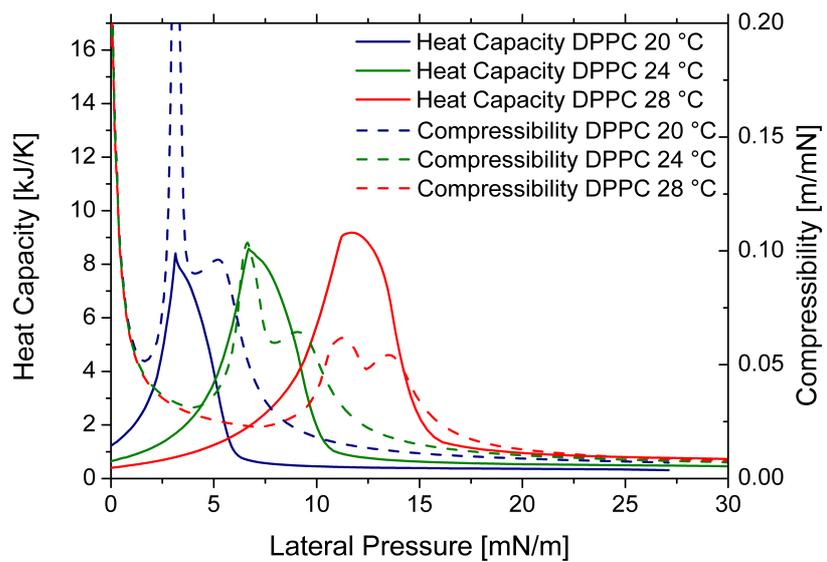

**Figure 5**

**Figure 1** The picture inset shows the system of springs connecting each lipid to a hexagonal lattice used for the model of the MC Simulations at hand. Every lipid has its own set of six springs, therefore creating connections over two springs to each of the six nearest neighbours. The spring constants $k_{f/g}$ and resting lengths $x_{0f/0g}$ of the disordered (*f*) and ordered (*g*) state are defined by the lipids compressibility $\kappa_{f/g}$ and area $A_{f/g}$ in the according state. For the simulations here, DPPC Monolayers are considered, so that the two isotherm and compressibility curves at *3°C* and *42°C* shown in the graph and its inset can be used to determine the compressibility $\kappa_{f/g}$ and area $A_{f/g}$ of the disordered (*f*) or ordered (*g*) state of a DPPC lipid. $A_{f/g}$ are defined by the first rise of the isotherm curve of the ordered (*3°C*) or the disordered (*48°C*) monolayer, $\kappa_{f/g}$ by the minima of the compressibility curves (calculated using the definition $\kappa_T = \frac{-1}{A}\left(\frac{\partial A}{\partial \Pi}\right)_T$ ).

**Figure 2** Simulation cycle used for the implementation of the model at hand. After a random lipid is selected two possible steps can be performed. A "classical" MC-step, where the lipid state may change according to the probability defined by the Gibbs free Energy change *ΔG*. Additionally a movement-step is possible, where the selected lipid is moved according to the forces defined by the springs attached to the lipid using a time constant $\delta$ and a friction constant $c_f$ following equ. (9).

**Figure 3 a)** Example for the determination of the integrated difference between a simulated and a measured isotherm at *24°C*. The integrated difference is defined by the shaded area in the graph and calculated for different combinations of $\delta$ and $c_f$. **b)** Course of values for the integrated differences for a row of $\delta$s for different valueas of $c_f$. Since the behaviour of the simulations at hand is not substantially depending on $\delta$ and $c_f$ (only small changes occur due to changes in $\delta$ or $c_f$), the choice of the latter is quite arbitrary. A short discussion in the text motivates the values of $\delta=2.4\cdot10^{-14}$ and $c_f=4\cdot10^{-15}$, which are eventually chosen.

**Figure 4 a)** Calculated (*continuous line*) and measured (*dashed line*) isotherms of DPPC Monolayers at three different temperature of *20°C*, *24°C* and *28°C* using a lattice size of *101x102* lipids at a runtime of *30K* simulation cycles per lipid. Calculation and simulation fit each other well, whereas the picture rows in **b)** show the corresponding screenshots of the monolayer lattice during phase transition as marked in a). The evolving ordered blotches may correspond to the according domains during phase transition observed by fluorescent or brewster angle microscopy in experiments. Comparable to experiments the simulations even show a general change of blotch shapes with the temperature.

**Figure 5** Heat capacity and compressibility curves calculated using the same simulations as for the isotherms and screenshots in fig. 4 at the three different temperature *20°C, 24°C* and *28°C*. Heat capacity and compressibility have their maximum at the same lateral pressures, although the detailed shapes differ. The compressibility curves have two maxima, which occur due to the split plateau of the isotherm shapes shown in fig. 4 (whereas the latter could be the effect of the finite lattice sizes used).